\documentclass[%
 reprint,
 amsmath,amssymb,
 aps,twocolumn,prl
]{revtex4-1}

\usepackage{graphicx}
\usepackage{hyperref}
\usepackage{multirow}
\usepackage{amsfonts}

\newcommand{\eqn}[1]{\begin{eqnarray} #1 \end{eqnarray}}
\newcommand{\tit}[1]{\textit{#1}}
\newcommand{\tbf}[1]{\textbf{#1}}
\newcommand{\trm}[1]{\textrm{#1}}

\newcommand{\tr}[1]{  \textrm{Tr}\left[ #1 \right]  }
\newcommand{\trc}[2]{  \textrm{Tr}_{#1}\left[ #2 \right]  }
\newcommand{\zum}[2]{\displaystyle\sum_{#1}^{#2}}

\newcommand{\ket}[1]{| #1 \rangle}
\newcommand{\ketbra}[2]{| #1 \rangle \langle #2 |}

\begin{document}

\title{A time-reversible quantum causal model}

\author{Jacques Pienaar}
\affiliation{
 International Institute of Physics, Universidade Federal do Rio Grande do Norte, Campus Universitario, Lagoa Nova, Natal-RN 59078-970, Brazil.
}

\date{\today}



\begin{abstract}
Modern approaches to causal modeling give a central role to interventions, which require the active input of an observer and introduces an explicit `causal arrow of time'. Causal models typically adopt a mechanistic interpretation, according to which the direction of the causal arrow is intrinsic to the process being studied. Here we investigate whether the direction of the causal arrow might be a contribution from the observer, rather than an intrinsic property of the process. Working within a counterfactual and non-mechanistic interpretation of causal modeling developed in Ref. \cite{LONGPAPER}, we propose a definition of a `quantum observational scheme' that we argue characterizes the observer-invariant properties of a causal model. By restricting to quantum processes that preserve the maximally mixed state (unbiasedness) we find that the statistics is symmetric under reversal of the time-ordering. The resulting model can therefore accommodate the idea that the causal arrow is observer-dependent, indicating a route towards reconciling the causal arrow with time-symmetric laws of physics.
\end{abstract}

\maketitle

\section{Introduction}

There are many physical phenomena whose dynamics are asymmetric along the time axis, providing `arrows of time' \cite{ZEH,PRICEBOOK,HALLIWELL}. Among the least studied of these is the \tit{causal} arrow of time: an action we take at time $t=T$ always seems to have consequences on one side of the co-ordinate time axis (that we label by convention with increasing times $t>T$), and never seems to have consequences on the other side of that axis (labeled with decreasing times $t<T$) \cite{PEARL,PRICEBOOK,COECKELAL,ORESH}. This asymmetry cannot be explained away by assuming that actions affecting the past would lead to logical paradoxes, for non-paradoxical accounts of backwards-in-time causation can readily be found in philosophy \cite{SEPtimetravel}, physics\cite{SEPtimemachines} and science fiction \footnote{The reader can probably fill in their own examples. My favourite is Shane Carruth's film, \tit{Primer}.}. The origin of the time asymmetry of causality remains an open question. \footnote{There are actually two aspects to this asymmetry: first, that actions always produce consequences in the same temporal direction for repeated actions on the same system; second, that the directions so obtained for any \tit{pair} of distinct systems are always found to be in agreement.}

This work aims to contribute to this big question by asking a smaller question: is the perceived direction of the causal influence intrinsic to the system, or is it a relative property of how the observer interacts with the system? The answer is not immediately obvious, because causal relations are ascertained by taking actions on a system and observing the results, which requires the participation of both the observer and the system. We emphasize that what is at stake is not a metaphysical question but a practical one: does the direction of causality \tit{necessarily} appear in an observer-independent description of a phenomenon? If not, then we can deduce that it is a property that depends in some way on the observer, either because it is a relative property of the observer and the phenomenon, or because it is entirely contributed by the observer. For example, the statue of Aristotle in Thessaloniki has a rest-mass which may be considered an intrinsic property of the statue, because its value is the same regardless of how and by whom it is measured. Its velocity is an observer-relative property because it cannot be specified except in relation to the velocity of the approaching tourist. Finally, the history of the statue is neither a property of the statue in itself nor a property of its physical relation to us, but rather is contained in records of our perceptions of the statue. It is natural to wonder: is the perceived direction of causality intrinsic like rest-mass, relative like velocity, or is it part of the memory and perceptual apparatus of the observer?
Progress on this question can only be made within a conceptual framework that gives `causality' a more precise meaning. One approach that has proven to be successful in a wide range of scientific disciplines is \tit{causal modeling} \cite{PEARL, SGS, WOODW}. In a causal model, there are essentially two modes by which an observer may ascertain causal relations. The first is called an \tit{observational scheme} and represents a form of data collection that is `passive' in some appropriate sense, usually understood to mean that the system is unaffected by the act of observation. The second mode of interaction is an \tit{interventionist scheme} in which the observer probes or disrupts specific variables of the system and observes the effect on the other variables. The conceptual difference between these two modes of interaction may be likened to the difference between observing a frog in a pond and dissecting it on a table: generally, watching the frog in the pond is sufficient to suggest some hypotheses about how it functions, but cutting it up is really the only way to know for sure. Perhaps because causal modeling has its roots in engineering and the practical sciences, \tit{causal models} strongly favour the view that causal relations, like the bones and musculature of the frog, are intrinsic to the system. Even if the direction of a cause cannot be ascertained from the \tit{observational scheme}, it is presumed to have some actual orientation that will be revealed once the observer makes an intervention.
Things become less clear when the systems under question are quantum. As we will see, the concept of an \tit{observational scheme} becomes essentially ill-defined for quantum systems. In the face of this, a few authors have proposed to re-define an \tit{observational scheme} in a way that remains true to the spirit of the classical definition \cite{RIED, KUEBLER}, however, these attempts face conceptual difficulties \cite{KUEBLER, LONGPAPER}. Many authors simply abandon the idea, choosing instead to interpret a \tit{quantum causal model} purely within the setting of an \tit{interventionist scheme} \cite{COSHRAP,ALLEN,KISS17}.
Any formulation of a causal model purely in terms of an \tit{interventionist scheme} denies us the means to separate the intrinsic from the observer-dependent aspects of causality. It is therefore necessary to propose a new definition of an \tit{observational scheme} suitable for quantum systems. This requires us to abandon any notion of causality that presupposes a \tit{mechanism}, and reinterpret causality as a kind of \tit{relation between counterfactuals}. For example, we can regard the causal structure as a relation that enables us to deduce, from the frog's movements in the pond, what would happen if we were to take it into the lab and probe each of its limbs separately. This view is compatible with a mechanical interpretation of causality, but does not require it.
In a longer companion work, a framework for causal modeling was developed based on this alternative interpretation of causality as relations among counterfactual experiments \cite{LONGPAPER}. The present work builds on that framework by using it to motivate a definition of a quantum observational scheme. It is shown that the probabilities obtained in the observational scheme plus the causal structure is sufficient to deduce what the probabilities would be for arbitrary interventions on any of the variables. In the special case of observations represented by `minimal' quantum instruments, like those used in Ref. \cite{LONGPAPER}, this result is found to hold if and only if the causal structure has the form of a `layered' DAG. This result is proven using the \tit{process matrix} definition of a quantum causal model given by other authors in Ref.\cite{COSHRAP}, restricted to the minimal `SIC-instruments' used in Ref. \cite{LONGPAPER}. I then ask what can be deduced from the quantum \tit{observational scheme} about the direction of causal influences in a system, without prior knowledge of the causal structure. It is found that when the processes in question are maximally-mixed-state preserving (unbiased), in stark contrast to the classical case, there is no information about the direction of causality available prior to intervention. This implies that causal modeling of quantum systems can be done in a way that does not \tit{require} a causal direction to be specified \tit{a priori}, as is usually assumed. It is therefore compatible with the notion that the direction of causal influence is an observer-dependent phenomenon, in line with the views expressed in eg. Refs. \cite{PRICEBOOK,ORESH,GUERIN}. 

\section{Functional models versus causal models}

To understand \tit{causal models} it is helpful to start with their conceptual predecessors: \tit{functional models}.\\
\tit{Definition:} A \tbf{functional model} is specified by a pair $\{G, \tbf{V} \}$ consisting of a directed acyclic graph (DAG) $G$ with nodes corresponding to $N$ variables $\tbf{V}=\{ V_i : i=1,2,...,N \}$; and a set of \tit{model parameters} $\{\eta_i, P_i(\eta_i), f_i : i=1,2,...,N\}$ where $\eta_i$ are independent random variables with probabilities $P_i(\eta_i)$, and $f_i$ are deterministic functions such that $V_i = f_i( \trm{pa}V_i, \eta_i )$ where $\trm{pa}V_i$ denote the parents of $V_i$ in $G$.\\
In a \tit{functional model}, probabilities for the variables $\tbf{V}$ in an \tit{observational scheme} are obtained by a simple procedure: one first samples a set of values from the $P_i(\eta_i)$ and then uses the functions $f_i( \trm{pa}V_i, \eta_i)$ to progressively determine the values of each $V_i$. The resulting distribution obeys a special constraint, called the \tbf{Causal Markov Condition} \cite{PEARL,SGS, HAUSWOOD}:
\eqn{ \label{eqn:CMC}
P_{\trm{\tbf{obs}}}(\tbf{V})= \prod^{N}_{i=1} \, P_{\trm{\tbf{obs}}}(V_i | \trm{pa}V_i) \, .
}
(Note that the form of this constraint depends on the sets $\trm{pa}V_i$ and hence is relative to the DAG $G$).
Conversely, if some $P(\tbf{V})$ satisfies \eqref{eqn:CMC} relative to a DAG $G$, then there exists a \tit{functional model} which generates $P(\tbf{V})$ in this way. Since \tit{functional models} treat the concept of `causal relation' as synonymous with `functional dependence', they also provide a direct way of formalizing the notion of \tit{intervention} on any variable $V_j$. Intuitively, an intervention is a physical action on a dependent variable $V_j$ in a system that cuts off this variable from its causes within the system, and allows it to be prepared in an arbitrary state by the experimenter. In a \tit{functional model}, this is done by introducing an experimentally controlled variable $X_j$ and replacing $f_j(\trm{pa}V_j,\eta_j)$ with the new dependence $f_j(X_j)$. The arrows connecting $\trm{pa}V_j$ to $V_j$ are then deleted from the DAG. The probability distribution generated from this intervened model also obeys a special constraint: 
\eqn{ \label{eqn:interv}
P_{\trm{\tbf{interv}}}(\tbf{V}|\trm{do}(V_j=x_j)) = \delta(V_j,x_j)\prod_{i \neq j} \, P_{\trm{\tbf{obs}}}(V_i | \trm{pa}V_i) \, , \nonumber \\
\, 
}
where $\delta(V_j,x_j) = 1$ if $V_j=x_j$ and $0$ otherwise, and the notation $\trm{do}(V_j=x_j)$ reminds us that we are forcing $V_j$ to the value $x_j$ by \tit{intervention}, instead of passively observing it. The collection of $N$ such distributions, representing interventions on each $V_i$, defines an \tit{interventionist scheme}. To make the further step to a \tit{causal model}, note that the form of both of the constraints \eqref{eqn:CMC}, \eqref{eqn:interv} is independent of the particular choices of \tit{model parameters} $\{\eta_i, P_i(\eta_i), f_i \}$ from which they were derived; in fact, it only depends on the causal structure $G$. Hence, only $G$ and $P_{\trm{\tbf{obs}}}$ are needed to describe observations at the empirical level. In the companion work \cite{LONGPAPER}, this is elevated to an axiom, called `causal sufficiency'. It is a key property of \tit{causal models} in our framework, because it allows us to elevate the concept of \tit{causal relation} to a universal status that transcends mere mechanism.\\ 
\tit{Definition:} A \tbf{causal model} is a pair $\{G, P_{\trm{\tbf{obs}}}(\tbf{V}) \}$, where $P_{\trm{\tbf{obs}}}(\tbf{V})$ satisfies the Causal Markov Condition \eqref{eqn:CMC} relative to $G$, and the results of interventions are given by the formula \eqref{eqn:interv}.\\
The importance of this step cannot be overstated. In a \tit{functional model}, causal structure was embodied in the deterministic mechanisms $f_i(\trm{pa}V_i,\eta_i)$, from which we derived the constraints \eqref{eqn:CMC}, \eqref{eqn:interv}. In a \tit{causal model}, as we interpret it here, this relationship is turned on its head: the constraints \eqref{eqn:CMC}, \eqref{eqn:interv} are taken as \tit{postulates} that serve to \tit{define} causal structure in terms of probabilities. This can be made more precise by noticing that the RHS of Eq. \eqref{eqn:interv} is entirely determined by the probabilities from the \tit{observational scheme}, plus causal structure. That is, `causal structure' merely codifies the precise way in which data obtained in the \tit{observational scheme} constrains the data that would be obtained for any conceivable intervention by an observer.

But what of the direction of causal influence? Curiously, the Causal Markov Condition \eqref{eqn:CMC} is not invariant under swapping the directions of all arrows in the DAG $G$. This asymmetry is usually understood in terms of it's simplest manifestation in the case of two nodes $V_2,V_3$ having a single parent $V_1$. In that case \eqref{eqn:CMC} is equivalent to the factorization of $V_2,V_3$ conditional on $V_1$, i.e. $P(V_2,V_3|V_1)=P(V_2|V_1)P(V_3|V_1)$, which is often associated with Reichenbach's Principle of Common Causes \cite{CAVLAL,SEPreich,REICHBOOK}. The asymmetry arises because reversing the direction of causal arrows in the DAG makes $V_1$ a common effect of $V_2,V_3$, and \eqref{eqn:CMC} does not require factorization of $V_2,V_3$ conditional on $V_1$ -- on the contrary, conditioning on a common effect of two variables typically correlates them, a result known in statistics as \tit{Berkson's effect} \cite{BERK}. This makes it possible in many cases to deduce the direction of causality purely from an \tit{observational scheme}. Traditionally, an \tit{observational scheme} is formalized as a method of making measurements that does not disturb the system. This allows us to interpret $P_{\trm{\tbf{obs}}}(\tbf{V})$ as representing information about the system that would be true even if \tit{nobody} had observed the system, i.e. information about the intrinsic properties of the system. It follows that the direction of causality is most naturally interpreted as one of the system's intrinsic properties.

\section{Quantum Interventions and Observations}

In the broadest sense, a \tit{quantum causal model} is any model that formalizes causal structure as a means of encoding the consequences of possible interventions on quantum systems. Unlike in the classical case, quantum systems have only been known to us under highly controlled laboratory settings. It is therefore natural that the first attempts at causal modeling of these systems have focused on finding a quantum generalization of the \tit{interventionist scheme}. These models can be summarized by an equation analogous to \eqref{eqn:interv} having the form of a `generalized Born rule' \cite{COSHRAP,OCB,ORESH} :
\eqn{ \label{eqn:qborn}
P(\tbf{V}|x_1,x_2,\dots x_N) = \tr{M^{x_1}_{v_1} \otimes \cdots \otimes M^{x_N}_{v_N} \cdot W  } \, . \nonumber \\
}
In this expression, each $M^{x_i}_{v_i}$ is the Choi-Jamio\l kowski matrix representation of a completely positive (CP) map $M^{x_i}_{v_i}(\rho): \mathcal{H}^{\trm{in}}_{V_i} \mapsto \mathcal{H}^{\trm{out}}_{V_i}$, which represents the effect of setting the control variable $X_i$ to the value $x_i$, and obtaining the outcome $v_i \in \trm{dom}(V_i|x_i)$ from the domain of possible values $\trm{dom}(V_i|x_i)$ (note that this domain might be different for different choices of $x_i$). The set of CP maps $\{ M^{x_i}_{v_i} : v_i \in \trm{dom}(V_i|x_i) \} := \mathcal{M}^{x_i}_{V_i}$ defines a \tit{quantum instrument}, which replaces the classical notion of \tit{intervention}, and $W$ is a positive semi-definite \tit{quantum process matrix} that maps arbitrary choices of quantum instruments to probability distributions over the observed outcomes $\tbf{V}$ \cite{COSHRAP}. Conceptually this model treats causality in much the same way as in a \tit{functional model}, with $W$ standing in the role of the \tit{model parameters} and where \tit{causal structure} refers to the structure of the \tit{quantum process} $W$ from which constraints on the probabilities are derived. This view implicitly supports interpreting $W$ (and hence the direction of causal influence) as a property intrinsic to the system being probed. At least, if one wishes to argue otherwise, it is necessary to supplement this description with an \tit{observational scheme} that would allow us to define $P_{\trm{\tbf{obs}}}(\tbf{V})$ as containing only the observer-independent information.

The usual definition of an \tit{observational scheme} in terms of non-disturbing measurements is problematic for quantum systems. On one hand, in order to perform inference, the measurements need to be informative about the system, but this implies that they disturb the state of the system. This feature of quantum systems has been formalized under the slogan ``no information without disturbance" \cite{BUSCH}.
Let us therefore re-evaluate the meaning of an \tit{observational scheme} by looking carefully at the role it actually plays in a \tit{causal model}. Two notable features emerge. First, from Eq. \eqref{eqn:CMC} we see that the scheme is \tit{passive} in the sense that it requires no active choices from the observer: there are no `control variables', there are only outcomes of a fixed set of measurements. Second, from Eq. \eqref{eqn:interv}, we see that the data from the \tit{observational scheme} plus the causal graph $G$ is sufficient to determine the results of an arbitrary intervention. This latter feature points to an interesting new way of thinking about an observational scheme: not as a scheme that reveals the system's properties in absence of observation (as usually assumed), but as a scheme that indicates the system's behaviour in the presence of \tit{any} observation. The information obtained in such a scheme depends on the \tit{fact} of observation, but not the particular features of the observation. It thus represents characteristics that may be called \tit{intrinsic} to the system, in the sense that they obtain independently of \tit{which} observation is made. Thus, what gives causal relations their `objective' character in an observational scheme is not that we are taking a \tit{view from nowhere}, but rather that we are adopting the \tit{view-point of no-one in particular} \cite{FINE}.
Existing proposals for quantum \tit{observational schemes} fail to meet both of the above requirements exemplified in Eqs. \eqref{eqn:CMC}, \eqref{eqn:interv}. The \tit{informationally symmetric} measurements proposed in Refs. \cite{RIED,RIEDPHD}, as well as the \tit{active quantum observational scheme} in Ref. \cite{KUEBLER} both allow the observer to actively choose which instruments to apply in each experimental run, hence are not \tit{passive}. Both definitions moreover have limited power to perform tomography of the \tit{quantum process} $W$, and hence cannot meet the second requirement of enabling one to deduce the results of an arbitrary intervention. We will now propose a definition of a \tit{quantum observational scheme} that is \tit{passive} in the sense described above, i.e. that does not depend on the active selection of control variables by an observer, and show that complete tomography is possible with such instruments. In order to give this definition, it is useful to introduce the concept of an \tit{informationally complete instrument}. To do so, we use the fact that we can express a CP map $M_{v_i}(\cdot)$ in terms of its Kraus decomposition:
\eqn{
M_{v_i}(\cdot) = \zum{k}{} \, A_{k,v_i} (\cdot)  A^{\dagger}_{k,v_i} \, .
}
In terms of this decomposition, for a given input state $\rho$ we may express the probability of outcome $V_i=v_i$ and the corresponding post-measurement state as (respectively):
\eqn{
P(V_i = v_i) &=& \tr{ \rho \left( \zum{k}{}  A^{\dagger}_{k,v_i} A_{k,v_i} \right)} := \tr{ \rho F_{v_i}} \, , \nonumber \\
\rho_{v_i} &=& \frac{1}{P(v_i)} \zum{k}{}  A_{k,v_i} (\rho)  A^{\dagger}_{k,v_i} \, .
}\\

Informational completeness of the instrument  $\mathcal{M}_{V_i} := \{ M_{v_i} : v_i \in \trm{dom}(V_i) \}$ can be expressed operationally by imagining a situation in which the instrument is applied to an unknown state $\rho$ and then the output state from the instrument is subjected to an unknown general measurement (POVM) $\{ B_i \}$. Informational completeness of $\mathcal{M}_{V_i}$ means that it must be possible to completely reconstruct both $\rho$ and $\{ B_i \}$ from the set of probabilities $P(V_i)$ of the instrument's outcomes. Formally, this is guaranteed by the following conditions:\\

\tit{Definition:} An \tit{informationally complete instrument} $\mathcal{M}_{V_i}$ is a quantum instrument with the properties that:\\
(i) The set of post-measurement states $\{ \rho_{v_i} : v_i \in \trm{dom}(V_i) \}$ are linearly independent and span $\mathcal{H}^{\trm{out}}_{V_i}$ ;\\
(ii) The set of POVM elements $\{ F_{v_i} : v_i \in \trm{dom}(V_i) \}$ are linearly independent and span $\mathcal{H}^{\trm{in}}_{V_i}$.\\

We are now ready to give our first general definition of a quantum observational scheme:\\

\tit{Definition:} A \tit{quantum observational scheme} assigns to each variable $V_i \in \trm{dom}(V_i)$ a single informationally complete instrument $\mathcal{M}_{V_i} := \{ M_{v_i} : v_i \in \trm{dom}(V_i) \}$, which is fixed for all runs of the experiment.\\

It is important to mention that an informationally complete instrument is not the same as an \tit{informationally complete set} (IC-set) of instruments, which is sometimes mentioned in the literature \cite{POLLOCKPRL,COSHRAP}. The latter refers to a \tit{set} of instruments $\{ \mathcal{M}^{x_i}_{V_i} : x_i \in \trm{dom}(X_i) \}$ targeted to a single variable $V_i$, which correspond to different possible settings of the control variable $X_i$ that may be varied in different experimental runs. The defining property of an IC-set is that when one applies such a set to every variable $V_i \in \tbf{V}$, then the resulting conditional probabilities $P(\tbf{V}|x_1,x_2,\dots x_N)$ can be used to perform complete tomography of the process $W$ in Eq. \eqref{eqn:qborn}. By contrast, an informationally complete instrument is a \tit{single} instrument that cannot be changed between experimental runs, and so a set of such instruments only produces a single set of `unconditioned' probabilities $P_{\trm{\tbf{obs}}}(\tbf{V})$. Moreover, perhaps counter-intuitively, these are in general not sufficient to perform complete tomography of $W$, despite the fact that each individual instrument is by itself informationally complete. This is essentially the same problem that arises in attempting to do tomography of non-Markovian processes; for more details see eg. Ref. \cite{MILZ16}.

Fortunately, it turns out that any IC-set of instruments for a given variable $V$ can be converted into a \tit{single} informationally complete instrument on $V$ \footnote{I am grateful to Kavan Modi for pointing this out.}. Formally, given an IC-set $\{ \mathcal{M}^{x}_{V} : x \in \trm{dom}(X) \}$, we can define CP maps
\eqn{
M_{(x,v)}(\cdot) := \gamma_{(x,v)} \, M^{x}_{v}(\cdot)
}
indexed by tuples $(X,V) := \{ (x,v) : x \in \trm{dom}(X), \, v \in \trm{dom}(V|x) \}$ where the coefficients $\gamma_{(x,v)}$ satisfy:
\eqn{
\zum{x,v}{} \, \gamma_{(x,v)} = 1 \, .
}
It is then straightforward to check that the set of these CP maps for all values of $(x,v)$ defines a single quantum instrument:
\eqn{
\mathcal{M}_{(X,V)} &:=& \{ M_{(x,v)}(\cdot) :   x \in \trm{dom}(X), \, v \in \trm{dom}(V|x) \} \, , \nonumber \\
&& \,
}
which is an informationally complete instrument for the variable $V$. Intuitively, we can imagine constructing this instrument from the IC-set by having, say, a machine automatically select the value of the control variable $X$ according to some sampling distribution determined by the coefficients $\gamma_{(x,v)}$.

Given an IC-set for each variable $V_i$, we can then use the above construction to define a single informationally complete measurement for each $V_i$, satisfying the requirements of a quantum observational scheme. Moreover, in this case, the statistics $P_{\trm{\tbf{obs}}}(\tbf{V})$ will be sufficient to perform complete tomography of $W$, by construction.

An interesting special case arises if we wish to insist that the informationally complete instruments are \tit{minimal}:\\

\tit{Definition:} An informationally complete instrument $\mathcal{M}_{V_i}$ is called \tit{minimal} if the input and output Hilbert space dimensions are the same, $\trm{dim}(\mathcal{H}^{\trm{in}}_{V_i})=\trm{dim}(\mathcal{H}^{\trm{out}}_{V_i}):=d_i$, and the number of possible values of $V_i$ is equal to exactly $d^2_i$.\\

One example of such a scheme was recently discussed in Ref. \cite{LONGPAPER}, where the instruments in question are \tit{symmetric informationally complete} (SIC) instruments. These are defined by:
\eqn{
M_{v_i}(\cdot) =  \frac{1}{d_i} \Pi_{v_i} (\cdot)  \Pi_{v_i} \, .
}
where $\{ \Pi_{v_i} : v_i \in \trm{dom}(V_i)$ are a set of $d^2_i$ rank-1 projectors satisfying:
\eqn{ \label{eqn:sics}
\tr{\Pi_{v_i} \Pi_{v_j}} = \frac{\delta_{ij}d_i + 1}{(d+1)} \, .
}
The usage of such instruments was motivated in Ref. \cite{LONGPAPER} by the desire to make a quantum causal model that is compatible with the `QBist' interpretation of quantum mechanics (see eg. Refs. \cite{NOTWITH,FAQBISM}). When the instruments in a quantum observational scheme are restricted to be SIC-instruments, we refer to this as a \tit{SIC-observational scheme}.

The restriction to instruments having only $d^2_i$ possible outcomes places severe constraints on what can be inferred about a general process matrix $W$ using only the statistics $P_{\trm{\tbf{obs}}}(\tbf{V})$. In fact, complete reconstruction of $W$ from the observed probabilities in a SIC-observational scheme will not be possible in general. It might be hoped, however, that it is possible for a suitably restricted class of processes $W$. Remarkably, we will show that complete tomography of $W$ is possible in a SIC-observational scheme if we merely restrict the \tit{causal structure} of the process:\\

\tbf{Result 1:} Tomography of a general process matrix $W$ compatible with the causal structure $G$ is possible in a SIC observational scheme if and only if $G$ is a \tit{layered} DAG (proof in Appendix A).\\

(\tit{Definition:} A DAG $G$ on a vertex set $\tbf{V}$ is said to be \tit{layered} if the nodes can be partitioned into $K$ subsets or `layers' $\tbf{V}=\tbf{L}_1 \cup \tbf{L}_2 \cup \cdots \cup \tbf{L}_K$ such that no member of a layer is a cause of any other member of the same layer, and such that for any triplet $\tbf{L}_i,\tbf{L}_j,\tbf{L}_k$ with $i<j<k$, each path connecting $\tbf{L}_i$ to $\tbf{L}_k$ is intercepted by $\tbf{L}_j$, i.e. contains a causal chain $A \rightarrow B \rightarrow C$ whose middle member $B$ is in $\tbf{L}_j$).\\

In order to uphold the key property of causal models -- that the \tit{observational scheme} should be sufficient to indicate what happens under an arbitrary \tit{intervention} -- it is therefore necessary and sufficient that the causal structure in a SIC-observational scheme should be \tit{layered} as described above. 

A further assumption that has been proposed in the literature is that $W$ should be an \tit{unbiased} process \cite{RIED,KUEBLER}, i.e. it should preserve the maximally mixed state. The standard reason for making this assumption is that it enables a more natural comparison between quantum and classical causal models when seeking to establish a quantum advantage for causal inference. However it also has an alternative motivation: in Ref. \cite{LONGPAPER} it was shown that the quantum generalization of the Causal Markov Condition for unbiased quantum processes is \tit{causally reversible}, that is, the condition is invariant under reversal of all arrows in the causal graph $G$. Here, we extend this result by proving the following:\\

\tbf{Result 2:} If the DAG of $W$ is layered \tit{and} the process $W$ is unbiased, then the probabilities $P_{\trm{\tbf{obs}}}(\tbf{V})$ obtained in a SIC-observational scheme are \tit{causally reversible}, that is, they are compatible with a quantum process whose DAG $G^*$ is obtained by reversing the directions of all arrows in $G$ (proof in Appendix B). \\

It is interesting to propose that the causal structure of quantum systems might be represented at the fundamental level by a SIC-observational scheme with a layered DAG. Under this assumption, it would follow that the direction of the causal arrows cannot be ascertained from $P_{\trm{\tbf{obs}}}(\tbf{V})$ alone. Thus, according to the arguments put forth in this paper, we should not regard this direction as an observer-invariant property of the system whose causal structure is described by the model. The direction of causality, we must suppose, is only revealed (or, perhaps more provocatively, produced) when an intervention is performed by a particular observer. This suggests that a time-symmetric description of the fundamental laws of physics may yet be compatible with the apparent asymmetry of the causal arrow of time.

It is natural to wonder what happens if one instead restricts all instruments to be projective measurements in a single basis, thereby obtaining the `classical limit' of a quantum causal model as defined in Ref. \cite{COSHRAP}. In this case, one would expect that the restriction to unbiased processes should lead to a causally reversible classical causal model. Such a model may represent a generalization of time-reversible Markov chains to arbitrary causal structures -- the investigation of this possibility is left to future work.

We remark in closing that the related work of Ref. \cite{ORESH} has shown that time-symmetric causality can be recovered by relaxing the constraints on quantum dynamics to allow for more general processes. Here, we have shown that symmetry can also be restored by \tit{restricting} quantum dynamics to \tit{unbiased} processes. In this sense, our work provides a complementary perspective on how time-symmetry of the causal arrow can be restored in quantum mechanics.

In conclusion, we have shown that it is possible to define a quantum observational scheme for arbitrary quantum processes, where the instruments are fixed for all runs of the experiment. More interestingly, we found that when the instruments are required to be minimal as in a SIC-observational scheme, the causal structure must be layered for complete tomography to be possible from the observed probabilities alone. Finally, we showed that under the restriction to unbiased processes, the model is causally reversible. Thus, the present work allows for the possibility that the observer has a fundamental role in determining the direction of causality in a system. The task of explaining exactly how this comes about remains an interesting open question. In particular, it would be interesting to investigate whether the causal arrow is grounded in purely \tit{physical} relations between observer and system, or whether it depends on more complex properties of the observer, such as their ability to perceive and store information.

\acknowledgments

I thank Kavan Modi and Fabio Costa for discussions that helped improve this work. This work was supported in part by the John E. Fetzer Memorial Trust.

\appendix

\section{Appendix A: Proof of Result 1}

Claim: Tomography of a general process matrix $W$ which is compatible with a given causal structure represented by a DAG $G$ is possible in a SIC- observational scheme if and only if $G$ is a \tit{layered} DAG. To prove this claim, we interpret \tit{compatible} to mean that $W$ \tit{factorizes} over the DAG $G$ as defined in Ref.  \cite{COSHRAP}. This will be explained below.

We first prove the `if' part. Let the nodes of $G$ be partitioned into $M$ sets $\tbf{Y}_1 \cup \tbf{Y}_2 \cup ... \cup \tbf{Y}_K$ corresponding to the layers of $G$ (we use `$\tbf{Y}$' here instead of $\tbf{L}$ because the lowercase `$\tbf{l}$' risks being confused with `1' and `I'). Let $\mathcal{H}^{I_j}_{\tbf{Y}_j}$ be the combined input Hilbert space of all nodes in the layer $\tbf{Y}_j$, and similarly let $\mathcal{H}^{O_j}_{\tbf{Y}_j}$ be the combined output Hilbert space from that layer. Then any process matrix $W$ that \tit{factorizes} over the DAG $G$ has the form \cite{COSHRAP}:
\eqn{ \label{eqn:divisible}
W=W^{I_1} \otimes W^{O_1I_2} \otimes ... \otimes W^{O_{K-1}I_{K}} \otimes \mathbb{I}^{O_K} \, ,
}
where $\mathbb{I}^{O_j}$ is the identity operator on $\mathcal{H}^{O_j}_{\tbf{Y}_j}$. The probabilities for a SIC-observational scheme are now obtained from the generalized Born rule \eqref{eqn:qborn}, by substituting in the SIC-instruments $\tilde{\mathcal{M}}_{V_i}$ for each node. Let us define $ \tilde{M}_{\tbf{y}_j}$ as the CP map on $\mathcal{H}^{I_j}_{\tbf{Y}_j}$ corresponding to the vector of values $\tbf{y}_j:= \{ v_i\} $ obtained for all nodes in the layer $Y_j$. Then the set of these maps for all sets of values $\tbf{y}_j$ defines a quantum instrument on the entire layer, $\tilde{\mathcal{M}}_{\tbf{Y}_j} := \{ \tilde{M}_{\tbf{y}_j} : \tbf{y}_j \in \trm{dom}(\tbf{Y}_j) \} $, which is also informationally complete. We can now write the observed probabilities from the SIC-observational scheme as:
\eqn{ \label{eqn:qpobs}
P_{\trm{\tbf{obs}}}(\tbf{y}_1, \tbf{y}_2,...,\tbf{y}_K) =  \tr{\tilde{M}_{\tbf{y}_1} \otimes \cdots \otimes \tilde{M}_{\tbf{y}_K} \cdot W  } \, . \nonumber \\
}

The definition of a SIC-instrument implies that the Choi-Jamio\l kowski matrices $\tilde{M}_{\tbf{y}_j}$ have the general form:
\eqn{
\tilde{M}_{\tbf{y}_j} = \Pi^{I_j}_{\tbf{y}_j} \otimes F^{O_j}_{\tbf{y}_j} \, ,
}
where 
\eqn{
\Pi_{\tbf{y}_j} &:=& \bigotimes_{v_i \in Y_j} \, \Pi_{v_i} \, \nonumber \\
F_{\tbf{y}_j} &:=& \beta_{\tbf{y}_j} \, \Pi_{\tbf{y}_j} \, \nonumber \\
\beta_{\tbf{y}_j} &:=& \prod_{v_i  \in Y_j} \, \beta_{v_i} \, .
}
In fact, for SIC-instruments we have $\beta_{v_i}:=\frac{1}{d_i}$ and the projectors $\Pi$ satisfy the overlaps in Eq. \eqref{eqn:sics}, however, these properties of SICs are not relevant to our present analysis. The present proof actually applies for \tit{any} quantum instruments that are minimal and whose POVM elements are rank-1 projections.

Continuing our analysis, note that the set of operators $\mathcal{F}_{\tbf{Y}_j}:=\{ F_{\tbf{y}_j} : \tbf{y}_j \in \tbf{Y}_j\}$ automatically forms an informationally-complete POVM for the whole layer. Inserting these operators and the factorized process from \eqref{eqn:divisible} into \eqref{eqn:qpobs}, we obtain:
\eqn{  \label{eqn:reconstruct}
P_{\trm{\tbf{obs}}}(\tbf{y}_1, \tbf{y}_2,...,\tbf{y}_K) &=& \tr{ F^{I_1}_{\tbf{y}_1} \cdot W^{I_1}_1 }  \nonumber \\
\times  \prod^{K}_{j=1} &&  \tr{\Pi^{O_j}_{\tbf{y}_j} \otimes F^{I_{j+1}}_{\tbf{y}_{j+1}}  \cdot W^{O_{j}I_{j+1}} } \nonumber \\
&=& P(\tbf{y}_1)P( \tbf{y}_2| \tbf{y}_1),...,P(\tbf{y}_K|\tbf{y}_{K-1}) \, , \nonumber \\
&&
}
where
\eqn{ \label{eqn:singleterm}
P( \tbf{y}_{j+1}| \tbf{y}_j) = \tr{ \Pi^{O_j}_{\tbf{y}_j} \otimes F^{I_{j+1}}_{\tbf{y}_{j+1}}  \cdot W^{O_{j}I_{j+1}}} 
}
is the conditional probability to obtain $\tbf{y}_{j+1}$ when measuring the POVM $\mathcal{F}_{\tbf{Y}_{j+1}}$ on the layer $\tbf{Y}_{j+1}$ given that the outcome $\tbf{y}_{j}$ was obtained on the previous layer $\tbf{Y}_{j}$. Since $\{ \Pi^{O_j}_{\tbf{y}_j} \otimes F^{I_{j+1}}_{\tbf{y}_{j+1}}  \}$ form a set of rank-1 projectors that spans the space of linear operators $\mathcal{L}(\mathcal{H}^{O_j}_{\tbf{Y}_j} \otimes \mathcal{H}^{I_{j+1}}_{\tbf{Y}_{j+1}})$, the probabilities $P( \tbf{y}_{j+1}| \tbf{y}_j)$ are sufficient to reconstruct an arbitrary sub-process $W^{O_{j}I_{j+1}}$, and hence using Eq. \eqref{eqn:reconstruct} we can reconstruct the full process $W$ from the observed probabilities $P_{\trm{\tbf{obs}}}(\tbf{V})$. \\
For the `only if' part of the proof, first note that for any DAG $G$ the nodes can be partitioned into sets $\{ \tbf{S}_j \}$ such that there is a directed path from a member of $\tbf{S}_i$ to $\tbf{S}_j$ whenever $j > i$, and no directed paths between any members of the same set. As with layers, one can define the rank-1 projectors $\Pi_{\tbf{s}_j}$ that together span the space of linear operators on the joint Hilbert space of all the nodes in the set $S_j$. Now, if $G$ is \tit{not} a layered DAG, then there must exist three sets $\tbf{S}_{j'}, \tbf{S}_{k'}, \tbf{S}_{l'}$ with $j'<k'<l'$ such that there is a path from $\tbf{S}_{j'}$ to $\tbf{S}_{l'}$ that does not intersect $\tbf{S}_{k'}$. Without loss of generality, we can choose the specific labels $j'=2,k'=3,l'=4$. Then the \tit{factorization} condition from Ref. \cite{COSHRAP} for the DAG implies that $W$ has the form:
\eqn{ \label{eqn:nondivisible}
W=W_{<2} \otimes W^{O_2 I_{3} O_{3} I_{4} } \otimes  W_{>4}\, ,
}
where by assumption $W^{O_2 I_{3} O_{3} I_{4} }$ cannot be further decomposed. Substituting this into \eqref{eqn:qpobs} we obtain a factorization of $P_{\trm{\tbf{obs}}}(\tbf{V})$ into a product of terms, which contains the term:
\eqn{
&& P(\tbf{s}_4 \ \tbf{s}_3|\tbf{s}_2) = \nonumber \\
&& \tr{\Pi^{O_2}_{\tbf{s}_2} \otimes F^{I_3}_{\tbf{s}_3}  \otimes \Pi^{O_3}_{\tbf{s}_3} \otimes F^{I_4}_{\tbf{s}_4} \cdot W^{O_2 I_{3} O_{3} I_{4} } } \, . \nonumber 
}
Now $W^{O_2 I_{3} O_{3} I_{4} }$ is a linear operator in the space $\mathcal{L}(\mathcal{H}^{O_2}_{\tbf{S}_2}\otimes \mathcal{H}^{I_3}_{\tbf{S}_3} \otimes \mathcal{H}^{O_3}_{\tbf{S}_3} \otimes \mathcal{H}^{I_4}_{\tbf{S}_4} )$, for which a spanning set of operators must have at least $(d_{2} d_{3}^2 d_{4})^2$ elements, with $d_{j}$ the dimension of the joint Hilbert space $\mathcal{H}_{\tbf{S}_j}$ associated to the set $\tbf{S}_j$. However, the set of projectors $\{ F_{\tbf{s}_2} \otimes F_{\tbf{s}_3}  \otimes F_{\tbf{s}_3} \otimes F_{\tbf{s}_4} \}$ for all sets of values of $\tbf{s}_{2},\tbf{s}_{3},\tbf{s}_{4}$ is required to have $(d_{2} d_{3} d_{4})^2$ elements in order to satisfy the requirement of being minimal, and so will in general not be sufficient to reconstruct the sub-process $W^{O_2 I_{3} O_{3} I_{4} } $ from the observed probabilities. This concludes the proof.

\section{Appendix B: Proof of Result 2}
Claim: If the DAG of $W$ is \tit{layered} and the process $W$ is \tit{unbiased}, then the probabilities $P_{\trm{\tbf{obs}}}(\tbf{V})$ obtained in a SIC-observational scheme are \tit{causally reversible}, that is, they are compatible with another quantum process $\bar{W}$ whose DAG $G^*$ is obtained by reversing the directions of all arrows in $G$. \\

To be more precise, our goal is to show that \tit{if} $P_{\trm{\tbf{obs}}}(\tbf{V})$ factorizes as shown in Eq. \eqref{eqn:reconstruct}, where each $W^{O_{j}I_{j+1}}$ is a valid unbiased quantum process representing a CPT map from from $\mathcal{H}^{O_j}_{\tbf{Y}_j}$ to $\mathcal{H}^{I_{j+1}}_{\tbf{Y}_{j+1}}$, \tit{then} $P_{\trm{\tbf{obs}}}(\tbf{V})$ must also factorize in the reverse order:
\eqn{ \label{eqn:reversedP}
P_{\trm{\tbf{obs}}}(\tbf{V}) = P( \tbf{y}_1| \tbf{y}_2),P( \tbf{y}_2| \tbf{y}_3),...,P(\tbf{y}_{K-1}|\tbf{y}_{K}) P(\tbf{y}_K) \, , \nonumber
}
such that each term can be expressed as:
\eqn{ \label{eqn:singlereverse}
P( \tbf{y}_{j}| \tbf{y}_{j+1}) = \tr{ \Pi^{O_{j+1}}_{\tbf{y}_{j+1}} \otimes F^{I_{j}}_{\tbf{y}_{j}}  \cdot \bar{W}^{O_{j+1}I_{j}}} \, , \nonumber 
}
where $\bar{W}^{O_{j+1}I_{j} }$ is a valid quantum process matrix representing a CPT map from from $\mathcal{H}^{O_{j+1}}_{\tbf{Y}_{j+1}}$ to $\mathcal{H}^{I_{j}}_{\tbf{Y}_{j}}$. First, note that the factorization of \eqref{eqn:reversedP} can be obtained directly from \eqref{eqn:reconstruct} by applying the standard formula for Bayesian inversion to each term:
\eqn{
P( \tbf{y}_{j+1}| \tbf{y}_{j}) = \frac{P( \tbf{y}_{j+1})}{P( \tbf{y}_{j})} \, P( \tbf{y}_{j}| \tbf{y}_{j+1}) \, . \nonumber 
}
Next, comparing \eqref{eqn:singlereverse} to \eqref{eqn:singleterm}, we see that the Claim is fulfilled only if we define
\eqn{ \label{eqn:barW}
\bar{W}^{O_{j+1}I_{j} } := \left\{ W^{O_{j}I_{j+1} } \right\}_{I \leftrightarrow O} \, \frac{P( \tbf{y}_{j})}{P( \tbf{y}_{j+1})} \frac{\beta_{j+1}}{\beta_{j}} \, ,
}
where $\{ \dots \}_{I \leftrightarrow O}$ means re-labeling the Hilbert spaces to switch their roles as `input' and `output' from their respective layers. It is important to note that a process which is a valid CPT map for one choice of input/output labelling will not in general be a valid CPT map when the input/output labels are switched: this is just the mathematical manifestation of the `causal arrow of time' for quantum processes\cite{COECKELAL,ORESH}. As a simple example, consider a quantum channel from the output of Alice's laboratory ($\mathcal{H}_{O_A}$) to the input of Bob's laboratory ($\mathcal{H}_{I_B}$) that discards Alice's output and produces a fixed pure state $\ket{\psi}$ at Bob's input. This is represented by the process matrix $W^{O_AI_B} = \mathbb{I}^{O_A} \otimes \ketbra{\psi}{\psi}^{I_B}$. Now consider the matrix obtained by switching the input and output:
\eqn{
\bar{W}^{I_AO_B} := \{ W^{O_AI_B} \}_{I \leftrightarrow O} =  \mathbb{I}^{I_A} \otimes \ketbra{\psi}{\psi}^{O_B}\, . \nonumber 
}
The Hilbert spaces are the same, but their roles are reversed: $\bar{W}^{I_AO_B} $ is now to be interpreted as a map from the \tit{output} of Bob's laboratory to the \tit{input} of Alice's laboratory. However, under this interpretation it does not represent a valid quantum process because
\eqn{
\trc{I_A}{\bar{W}^{O_BI_A}}= \ketbra{\psi}{\psi}^{O_B} \neq \mathbb{I}^{O_B} \, , \nonumber \\
\tr{\bar{W}^{O_BI_A}}= d_{I_A} \neq d_{O_B} \, .
}
Qualitatively, the first property means the map \tit{forces} the output of Bob's lab to be $\ketbra{\psi}{\psi}^{O_B}$ (regardless of Bob's efforts to produce something else) and hence represents a case of `deterministic post-selection' of Bob's output, which is not allowed by quantum theory. (Indeed, the assumption that there can be no deterministic post-selection is often called \tit{causality} in the literature \cite{CDP,ORESH}). The second property shows that the state produced at Alice's input is $\mathbb{I}^{I_A}$, which is not normalized. More generally, the conditions for any process matrix to represent a valid CPT map from $\mathcal{H}^{O_j}_{\tbf{Y}_j}$ to $\mathcal{H}^{I_{j+1}}_{\tbf{Y}_{j+1}}$ are given by \cite{OCB}:
\eqn{
W^{O_jI_{j+1}} &\geq& 0 \, , \nonumber \\
\trc{I_{j+1}}{W^{O_jI_{j+1}} } &=& \mathbb{I}^{O_j} \, , \nonumber \\
\tr{W^{O_jI_{j+1}}} &=& d_{O_j} \, .
}
Notice that these conditions are not symmetric under a re-labelling of $O \leftrightarrow I$, and hence it matters which Hilbert space is interpreted as the `output' of the preceding layer (hence the input to $W$) and which is the `input' to the next layer (hence the output from $W$). It is therefore important to verify whether the matrix $\bar{W}^{O_{j+1}I_{j}}$ defined in \eqref{eqn:barW} represents a valid quantum process. As we now show, this is guaranteed if the original process $W^{O_{j}I_{j+1}}$ is unbiased, which means it satisfies the additional constraint:
\eqn{ \label{eqn:unitality}
\trc{O_{j}}{W^{O_jI_{j+1}} } &=& \frac{d_j}{d_{j+1}}\mathbb{I}^{I_{j+1}} \, .
}
The conditions for $\bar{W}^{O_{j+1}I_{j}}$ to be a valid CPT map from $\mathcal{H}^{O_{j+1}}_{\tbf{Y}_{j+1}}$ to $\mathcal{H}^{I_{j}}_{\tbf{Y}_{j}}$ are given by:
\eqn{ \label{eqn:barconditions}
\bar{W}^{O_{j+1}I_j}  &\geq & 0 \, , \nonumber \\
\trc{I_{j}}{\bar{W}^{O_{j+1}I_j} } &=& \mathbb{I}^{O_{j+1}} \, , \nonumber \\
\tr{\bar{W}^{O_{j+1}I_{j}} } &=& d_{O_{j+1}} \, .
}
To prove that these conditions are met, we first simplify the expression \eqref{eqn:barW} using the fact that $W^{O_{j}I_{j+1} }$ is unbiased. Since the initial term $W^{I_1}_1$ of \eqref{eqn:divisible} can be thought of as a process from a trivial Hilbert space (with dimension $d=1$) to the space $\mathcal{H}^{I_{1}}_{\tbf{Y}_{1}}$, the condition \eqref{eqn:unitality} implies that $W^{I_1}_1 = \frac{1}{d_1} \mathbb{I}^{I_1}$. Next, we note that if one marginalizes over the outcome $v_i$ of an SIC-instrument (actually any minimal instrument with projective elements), the result is a CPT map that is automatically unbiased:
\eqn{
\zum{v_i}{} \, \tilde{M}_{v_i}(\mathbb{I}) &=& \zum{v_i}{} \,  \beta_{v_i} \Pi_{v_i} \, (\mathbb{I}) \, \Pi_{v_i} \nonumber \\
&=& \zum{v_i}{} \,  \beta_{v_i} \Pi_{v_i} = \mathbb{I} \, .
}
The above two results imply that for an unbiased process $W$ with a layered DAG, the marginal probability for the outcome set $\tbf{y}_j$ in a given layer $\tbf{Y}_j$ (i.e. after summing over the outcomes obtained in all other layers) is equal to the probability of obtaining $\tbf{y}_j$ when measuring the maximally mixed state, that is:
\eqn{
P(\tbf{y}_j) &=& \tr{\frac{1}{d_j} \mathbb{I} \cdot F_{\tbf{y}_j} } = \frac{\beta_j}{d_j} \, .
}
Substituting this into \eqref{eqn:barW} we obtain the simpler expression:
\eqn{ \label{eqn:barW2}
\bar{W}^{O_{j+1}I_{j}} := \left\{ W^{O_{j}I_{j+1}} \right\}_{I \leftrightarrow O} \, \frac{d_{j+1}}{d_{j}} \, .
}
Since $\frac{d_{j+1}}{d_{j}} > 0$, the first condition of \eqref{eqn:barconditions} is immediately met. For the second condition, with the help of \eqref{eqn:unitality}, we find:
\eqn{
\trc{I_{j}}{\bar{W}^{O_{j+1}I_j} } &=& \trc{I_{j}}{\left\{ W^{O_{j}I_{j+1}} \right\}_{I \leftrightarrow O}  } \, \frac{d_{j+1}}{d_{j}} \nonumber \\
&=& \left\{ \trc{O_{j}}{ W^{O_{j}I_{j+1}}  } \right\}_{I \leftrightarrow O}  \, \frac{d_{j+1}}{d_{j}} \nonumber \\
&=& \left\{  \frac{d_j}{d_{j+1}}\mathbb{I}^{I_{j+1}} \right\}_{I \leftrightarrow O}  \, \frac{d_{j+1}}{d_{j}} \nonumber \\
&=& \mathbb{I}^{O_{j+1}} \, ,
}
and
\eqn{
\tr{\bar{W}^{O_{j+1}I_{j} }}&=&  \left\{ \tr{ W^{O_{j}I_{j+1}} } \right\}_{I \leftrightarrow O} \, \frac{d_{j+1}}{d_{j}} \nonumber \\
&=& (d_j) \frac{d_{j+1}}{d_{j}} \nonumber \\
&=& d_{j+1} 
}
and hence all conditions \eqref{eqn:barconditions} are met. We conclude that $\bar{W}^{O_{j+1}I_j}$ is a valid process representing a CPT map from $\mathcal{H}^{O_{j+1}}_{\tbf{Y}_{j+1}}$ to $\mathcal{H}^{I_{j}}_{\tbf{Y}_{j}}$. Since the complete reverse process formed by
\eqn{
\bar{W}:=\mathbb{I}^{O_1} \otimes \bar{W}^{O_{2}I_{1}} \otimes ... \otimes \bar{W}^{O_{K}I_{K-1}} \otimes \mathbb{I}^{I_K} \, ,
}
clearly factorizes over the reversed DAG $G^*$, this completes the proof.
\end{document}